# Quantum-Correlation-Based Multi-Emitter Localization in Information-Limited Regimes via Learned Inference

**Sanket K. Tripathy [1,2,5] and Simeon I. Bogdanov [3,4,5]**

[1]Department of Physics, Indian Institute of Technology Madras, Chennai, India

[2]Department of Physics, The Grainger College of Engineering, University of Illinois Urbana-Champaign, Urbana, IL, USA

[3]Department of Electrical and Computer Engineering, The Grainger College of Engineering, University of Illinois Urbana-Champaign, Urbana, IL, USA

[4]Nick Holonyak, Jr. Micro and Nanotechnology Laboratory, University of Illinois Urbana-Champaign, Urbana, IL, USA

[5]Illinois Quantum Information Science and Technology Center, University of Illinois Urbana-Champaign, Urbana, IL, USA

**Abstract:**

Quantum correlation measurements have emerged as a powerful tool for localizing multiple emitters with subwavelength separation. Multi-emitter localization consists of two distinct tasks: identifying the emitter count $N$ and estimating their positions $\{x^i\}$. Existing fit-and-select frameworks fit the measured intensity and autocorrelation data using several models corresponding to different values of $N$ and then select the best-supported model. This procedure lets errors from the multi-parameter optimization of emitter positions propagate into the choice of model, degrading performance under photon-limited and high-background conditions. We propose a method that directly infers the emitter count and positions from the multimodal data via learned inference. Learned inference improves both the emitter-counting accuracy and localization precision compared to the fit-and-select method. It achieves more than tenfold improvement in precision and requires up to five times fewer photons to achieve practically meaningful localization with the largest gains observed under low photon counts, strong background, narrowly spaced emitter configurations, and spatially truncated data. This photon-efficient quantum-assisted localization framework offers a promising route toward real-time biological imaging and single-particle tracking in densely labeled samples.

**Introduction:**

Emitter localization lies at the heart of modern single-molecule microscopy and biological imaging [1], [2], [3]. As biological and nanoscale systems increasingly demand resolution beyond the diffraction limit, traditional intensity-based localization algorithms face critical limitations. In particular, limited photon budgets, background fluorescence and detector noise can substantially degrade emitter localization [4], [5], [6], [7], [8]. Overlapping point spread functions (PSFs) in densely labeled samples make the coordinate reconstruction especially challenging [9], [10], [11], [12], [13]. These factors motivate the development of localization methods that remain robust under a variety of adverse conditions encountered in practical imaging [14], [15], [16].

Photon correlation measurements involving the second-order autocorrelation function $g^{(2)}(\tau)$, offer a compelling quantum tool for overcoming these challenges. By leveraging the photon antibunching behavior intrinsic to single quantum emitters [17], [18], [19], photon correlation-based methods offer spatial resolution beyond the classical diffraction limit and help distinguish emitters from classical background sources. Recent studies have shown that spatially resolved intensity and $g^{(2)}(0)$ signals complement each other to achieve sub-diffraction multi-emitter localization, notably under adverse background and noise conditions [20], [21], [22].

Multi-emitter localization requires solving two related but distinct inference problems: determining how many emitters are present in the cluster (emitter count $N$) [23] and estimating their positions ($\{x^i\}_{i=1}^N$). A key feature of this inference problem is that the estimation of emitter positions is conditioned on the correct choice of *N*. The choice of *N*, in turn, changes the model's dimensionality. Existing approaches address this by fitting the measurements using several models corresponding to different values of *N* and then choosing the best-fitting model (fit-and-select approach) [24]. For example, the Akaike Information Criterion (AIC) is a logical choice for balancing spatial fit quality against model complexity [21], [25].

However, this method creates a strict dependency in which errors in the multi-parameter spatial fit propagate directly into the estimation of *N*. This dependency becomes fragile in photon-limited or high-background regimes, when emitter configurations with different *N* produce similar-looking data. Then, even modest fit errors in the parametric optimizer can lead to a global error, i.e a wrong model choice.

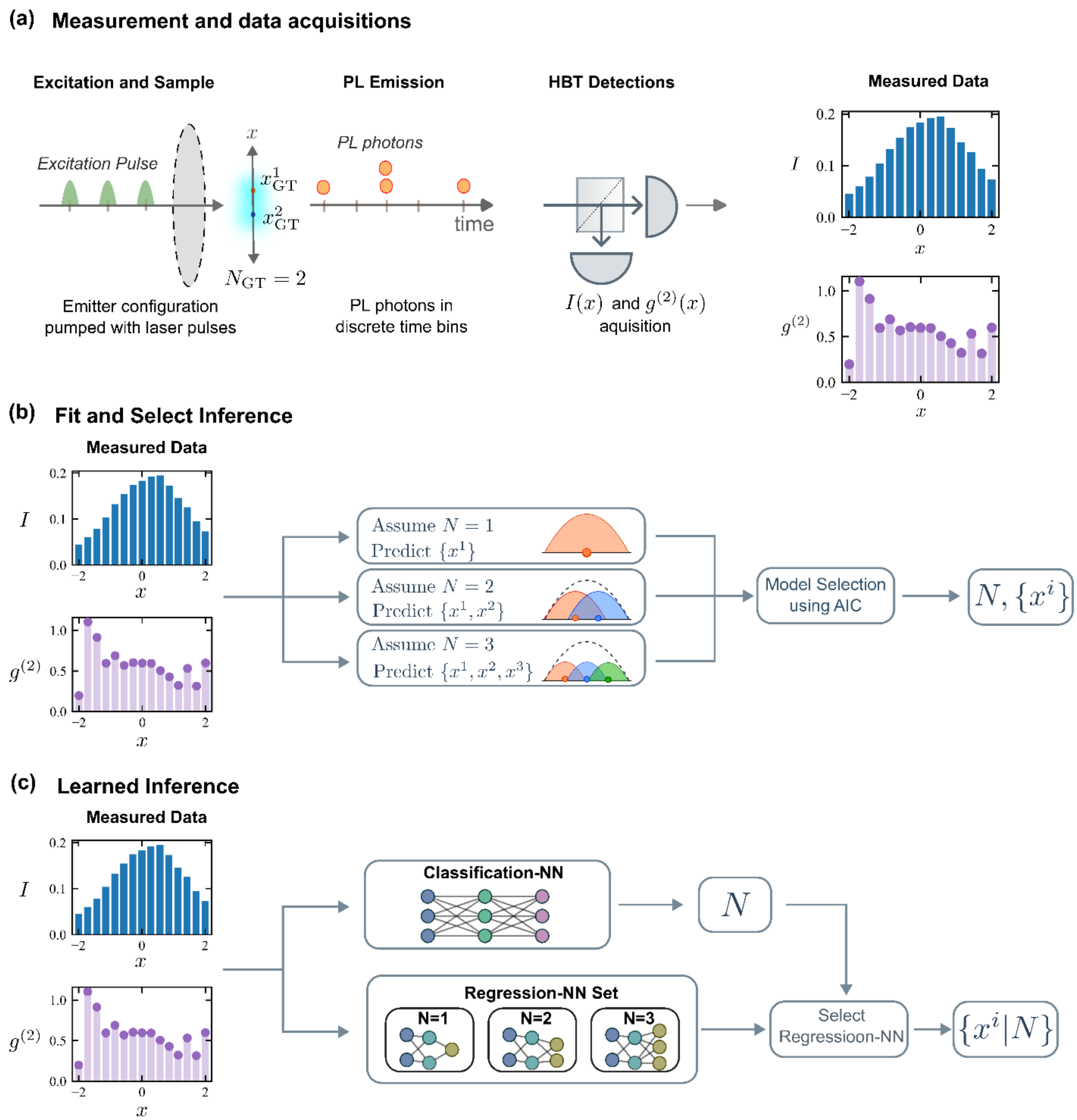

**Figure 1. Overview of the localization framework.**

(a) Multiple closely spaced emitters are excited by a pulsed pump. The photoluminescence (PL) is measured with a Hanbury Brown–Twiss (HBT) interferometer to obtain spatially sampled intensity *I*(*x*) and second-order correlation, $g^{(2)}(x)$, data. Representative measurements are shown on the right. (b) In the fit-and-select approach, intensity and $g^{(2)}(x)$ maps are fitted for each candidate emitter-number model using a weighted least-squares objective. AIC then selects the best-supported model, yielding a guess of *N* and $\{x^i\}$. (c) In the proposed learned inference approach, *I*(*x*) and $g^{(2)}(x)$ maps are directly used for both emitter-counting and position-estimation. The output of the Classification-NN (*N*) is used to select the correct Regression-NN and that yields an estimate of *N* emitter positions.

These limitations motivate the question of whether $N$ and $\{x^i\}$ can be inferred directly from the intensity and autocorrelation data, rather than determining $N$ through a post-hoc arbitration among competing spatial models [21], [26], [27]. In the next section, we show that in the fit-and-select approach, inference of $N$ can fail even when the position estimator remains precise in the model corresponding to the correct $N$. We also show that AIC's tendency to favor the lower-complexity model introduces a biased error in $N$. In our proposed method, one dedicated neural network (NN) infers the emitter count (the Classification-NN), while a second network infers emitter positions for a specified value of $N$ (the Regression-NN). In the results sections, we show that the Classification-NN substantially improves the emitter-counting accuracy and is significantly more robust to data degradation under vanishing emitter separation, low photon budgets and low signal-to-background ratios (SBR). Next, we show that based on a correct input for $N$, the Regression-NN also outperforms parametric fitting of emitter positions. Finally, we examine the robustness of direct learned inference to the spatial truncation of data when the emitters are positioned close to the edge of the measurement volume.

**Problem Statement**.

We begin by briefly defining the setup of the simulation. The technical details of this setup are available in Appendix A. We consider a scenario (schematically represented in Fig. 1(a)) where a collection of $N \in \{1,2,3\}$ single-photon emitters (SPEs) is located within the focal volume of a scanning microscope objective at the ground truth coordinates $x^i_{\mathrm{GT}}$. These SPEs are subject to a sequence of $n_{\mathrm{pulse}}$ optical pump pulses delivered through a scanning objective. $n_{\mathrm{pulse}}$ defines the photon budget for the simulation. The SPEs have a Gaussian PSF with a given typical width $\sigma$ corresponding to the diffraction limit of the optical system and taken as the unit of spatial scale in this paper. Each SPE emits a perfectly antibunched photoluminescence with peak emission probability per pulse $A^i_{\mathrm{GT}}$ on top of a classical background. The background is treated as a source with uniform intensity $I_{\mathrm{BG}}$ i.e., a Poisson random variable with a known average. The emission is collected by the objective and routed to a Hanbury–Brown–Twiss interferometer. The objective is scanned over the vicinity of the confocal volume, producing local $I(x)$ and $g^{(2)}(\tau = 0, x) \equiv g^{(2)}(x)$ maps with $n_{\mathrm{pulse}}$ pulses per pixel. We model the confocal volume in 1D as the segment $[-2,2]$ with samples collected over a uniform grid of 15 pixels. The intensity profile $I(x)$ is the average number of clicks per pulse from both detectors, calculated over $n_{\mathrm{pulse}}$ pulses, which explicitly depends on the system's excitation, emission, and detection efficiencies. $g^{(2)}(x)$ is defined as the ratio between the number of coincident (zero-delay) clicks over the average number of non-coincident clicks per time bin over the same period. The signal-to-background ratio (SBR) is defined as the ratio between the global maximum of $I(x)$ and the background emission intensity $I_{\mathrm{BG}}$.

Given the $I$ and $g^{(2)}$ maps, localization requires (1) estimating the emitter count $N$ in the focal volume, (2) estimating the SPE positions conditioned on the emitter count, i.e., $\{x^i|N\}_{1\leq i\leq N}$. In the fit-and-select approach, these two steps are linked because $N$ is not directly inferred from the measurements. In this approach, one first estimates the continuous parameters associated with each candidate model, i.e., $\{x^i, A^i|N\}_{N\in\mathbb{N}}$ and only then decides which model best fits the data. Importantly, even though position estimation may be relatively precise, the emitter count inference may fail due to two separate effects. First, as photon noise, background, or spatial truncation make the underlying parameter fits increasingly ambiguous, uncertainty in those fits propagates into the subsequent model-selection step. Moreover, when the method misclassifies two-emitter data, the errors are strongly asymmetric. It preferentially selects the simpler $N$ = 1 model rather than the more complex three-emitter model. The improvement in fit provided by introducing an additional emitter is not sufficient to offset the AIC complexity penalty. Figure 2 demonstrates this behavior.

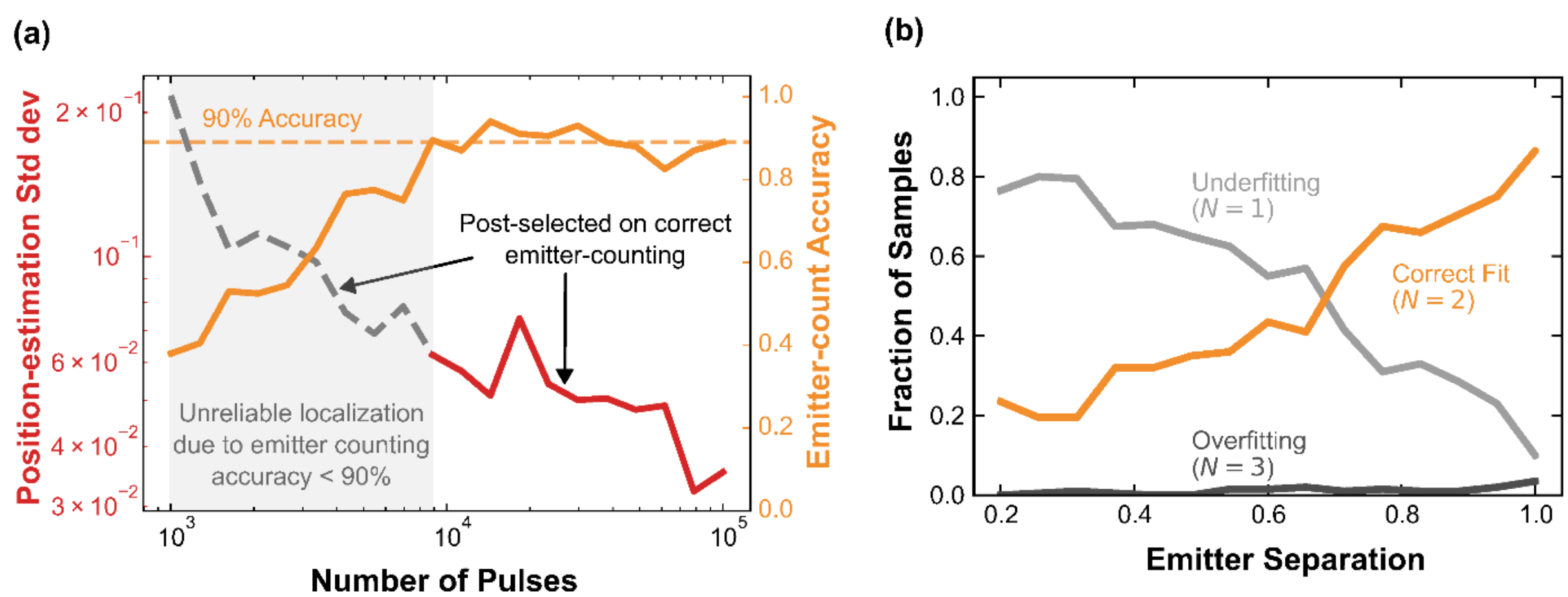

**Figure 2. The limitations of the fit-and-select approach.**

(a) Position-estimation precision and emitter-counting accuracy as a function of the photon budget $n_{\mathrm{pulse}}$ over a representative set of $N_{\mathrm{GT}} = 2$ emitter configurations. The position-estimation standard deviation (red, left axis) is calculated only when the guess of $N$ is correct, while the emitter-counting accuracy (orange, right axis) quantifies the probability of identifying the correct $N$. The dashed horizontal line indicates 90% emitter-counting accuracy. (b) Fraction of samples classified as $N = 1$, 2 or 3 emitters as a function of the emitter separation for $N_{\mathrm{GT}} = 2$.

In Fig. 2(a), we plot the simulation results for $N_{\mathrm{GT}} = 2$ non-identical emitters with fixed ground truth separation $|x_{\mathrm{GT}}^1 - x_{\mathrm{GT}}^2| = 0.5$. The emitter brightnesses are chosen to produce a fixed SBR of 10 at the peak of the total intensity profile. The detailed weighted least-squares fitting, AIC formulation, and model-selection procedure are provided in Appendix B. The position-estimation precision increases continuously with increasing photon budgets. However, the fit-and-select approach's emitter-counting accuracy deteriorates substantially below for $n_{\text{pulse}} < 10^4$. This happens even though the position-estimation precision of the $N = 2$ model deteriorates only modestly. In Fig. 2(b), we plot the predicted emitter-count vs increasing emitter separation for a case with $N_{\text{GT}} = 2$, SBR =10. The method is seen to consistently underfit, choosing $N = 1$ for lower emitter separation where fitting is unable to discern the two emitters, while $N = 3$ is almost never chosen. This shows that the fit-and-select approach also suffers from a bias towards simpler models. A reliable end-to-end localization can only occur if the value of $N$ is accurately identified. Therefore, model selection becomes the limiting component of the localization pipeline.

**Proposed Localization Pipeline: Learned Inference**

To counter the vulnerabilities of the fit-and-select approach, we introduce another emitter localization pipeline based on separate neural networks, specializing in either inferring the emitter count or the emitter positions. In the following sections, we examine emitter counting and position estimation separately, allowing us to determine where learned inference provides an advantage in each part of the localization problem.

We first consider the inference of $N$ as a standalone task. In the fit-and-select approach, the emitter counting is coupled to the estimation of emitter positions and emission probabilities through the fitting of competing models. To decouple counting from these spatial parameters, a specialized neural network, hereafter referred to as the Classification-NN, predicts $N$ directly from $I(x)$ and $g^{(2)}(x)$.

For the second inference task of position estimation conditioned upon the estimation of $N$, in principle, the weighted least-squares estimator from the fit-and-select approach can be reused. However, the statistically optimal fusion of $I(x)$ and $g^{(2)}(x)$ data for each task is nontrivial. In particular, the probability distributions of the derived $I(x)$ and $g^{(2)}(x)$ observables do not generally admit convenient closed-form expressions, and previous likelihood-based approaches have therefore relied on approximations, for example, approximating the distributions of both observables as Gaussian in constructing their maximum-likelihood estimator [26], [27]. Existing few-emitter approaches instead combine intensity and correlation residuals through a prescribed weighted least-squares objective. The optimal relative weighting of the two observables has been shown to depend on measurement conditions, including background [21], [28]. In view of these difficulties in defining the likelihood function, we additionally consider a data-driven regression estimator that learns the inverse mapping from the measured $I$ and $g^{(2)}$ profiles directly, allowing nonlinear and condition-dependent use of the two measurement modalities without requiring an explicit joint likelihood or a fixed scalar weighting. We find that a dedicated set of neural networks, Regression-NNs, each trained for a specific value of $N$ outperforms the corresponding weighted least-squares optimizers.

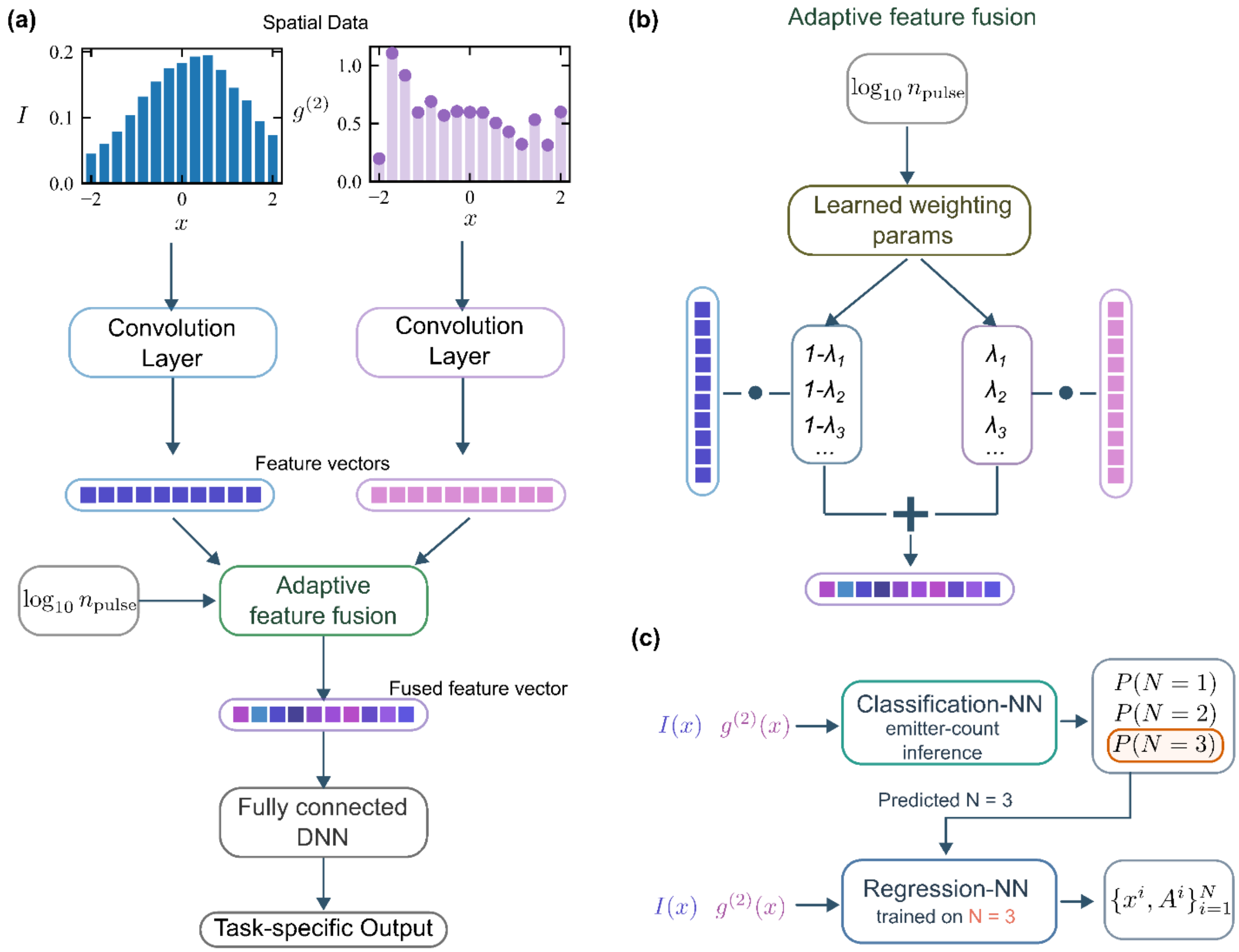

**Figure 3. Architecture and learned inference pipeline of the proposed localization framework.** (a) General architecture shared by the classification and regression models. The measured intensity, *I(x)*, and second-order correlation, $g^{(2)}(x)$, are processed by separate convolutional feature-extraction branches. The resulting feature vectors are combined by an adaptive feature-fusion layer using photon-budget-dependent weighting and subsequently passed to a fully connected task-specific network. (b) Schematic of the adaptive feature-fusion operation, in which feature vectors from the intensity and autocorrelation branches are weighted independently and combined to form a fused feature representation. (c) Learned inference procedure. In the first stage, the Classification-NN maps *I(x)* and $g^{(2)}(x)$ to probabilities for the possible emitter numbers, from which the highest-probability hypothesis $N$ is selected. In the second stage, the regression model corresponding to the selected $N$ is applied to the same measurements to estimate the emitter positions and emission probabilities.

Both of the above-mentioned NNs use a dual-branch convolutional architecture in which $I(x)$ and $g^{(2)}(x)$ maps are processed by separate feature-extraction branches that learn task-specific representations of the underlying optical signal (Fig. 3(a)). The resulting feature vectors are combined through an adaptive feature-fusion layer whose weighting depends on the number of excitation pulses. The fused representation is subsequently passed to a fully connected task-specific network (Fig. 3(b)).

The Classification-NN outputs a probability distribution over the possible emitter counts and the value *N* with the highest predicted probability is selected. The Regression-NN, trained specifically on data with *N* emitters, is then employed to estimate the emitter positions (Fig. 3(c)). The following sections describe the classification and position-estimation networks in detail and evaluate their performance independently.

**Results: Emitter Counting (Classification-NN)**

We treat the task of emitter counting as a supervised statistical inference problem [29], [30], [31], where $I(x)$ and $g^{(2)}(x)$ are mapped directly to discrete labels $N$. We employ a dual-branch convolutional neural network (CNN) that takes as input $I(x)$ and $g^{(2)}(x)$ measurements (Fig. 3). The two input channels are processed separately by identical CNNs, each consisting of two one-dimensional convolutional layers followed by global average pooling, yielding feature vectors $f_I, f_{g2}$. To account for varying photon budgets across samples, the logarithm of the number of excitation pulses $\log_{10}(n_{\text{pulse}})$ is provided as an auxiliary input and transformed into a gating vector $\bar{\lambda}$. This gating adaptively weights the relative contribution of intensity and autocorrelation features elementwise according to

$$f = (1 - \bar{\lambda}). f_I + \bar{\lambda}. f_{g2} \quad (1)$$

allowing the network to combine autocorrelation and intensity features flexibly depending on the photon budget. The fused representation is subsequently passed to a fully connected network, yielding a probability distribution over *N*. The highest probability outcome is chosen as the guess for the number of emitters. For more details on the network and their training, please refer to Appendix C.

To benchmark the emitter-counting performance, we directly compare the Classification-NN with the fit-and-select approach, evaluating both methods on identical simulated datasets and using emitter-counting accuracy as the primary metric. We run both methods on $N_{\text{GT}}$ = 2, 3 non-identical SPE configurations with randomized positions. N = 1 is a trivial case, and both methods achieve near-perfect accuracy, and its results are not plotted. We perform Monte Carlo simulations and analyze the accuracy as a function of the minimum emitter separation ($\min_{i \neq j} |x_{\text{GT}}^i - x_{\text{GT}}^j|$) and $n_{\text{pulse}}$. The emitter separation was sampled uniformly in the range $[0.1, 1.0]$. For analysis, samples were grouped into bins of width 0.1 based on the ground-truth separation, e.g., separations in $[0.1, 0.2)$, $[0.2, 0.3)$, etc. are labelled with the mid-point of the range. Each bin had 100 samples for each value of $n_{\text{pulse}}$. We choose values of SBR = 2, 6, and 10 to span a range of conditions with strong background.

Fig. 4 shows the comparison of emitter-counting performance between the Classification-NN and the fit-and-select approach. For SBR = 6 and 10, the NN greatly outperforms the fit-and-select approach. The NN maintains an accuracy of 70–90% across the swept parameter regime, whereas the fit-and-select approach's accuracy degrades rapidly at lower photon budgets and smaller separations, reaching well below 50%, corresponding to a random-choice accuracy. This shows that the NN approach is significantly more robust to information degradation.

For $N_{\text{GT}} = 2$ and SBR=2, that the NN has only a slight advantage over the fit-and-select approach although both methods are impractical due to low accuracies. For $N_{\text{GT}} = 3$ and SBR=2, the NN achieves an average accuracy of ~60%, and up to >80% at higher emitter separations. The fit-and-select approach shows accuracy consistently around 10%.

**Results: Position Estimation (Regression-NN)**

We have established that the Classification-NN reliably determines $N$. Now, we consider the second stage of the localization pipeline i.e., estimating emitter positions, assuming $N$ is known. We compare two methods: position estimation using the fit-and-select approach and a dedicated Regression-NN. We evaluate the performance assuming prior successful counting, i.e., $N_{\text{GT}}$ is known, thereby isolating the regression problem from counting errors. This allows us to directly compare the intrinsic position-estimation capability of the proposed NN against the fit-and-select approach, in which weighted-least-squares fitting is used for position estimation.

The Regression-NN uses a dual-branch CNN similar in spirit to the Classification-NN (Fig. 3), but with a few architectural modifications motivated by the fundamentally different nature of the task. Whereas emitter-counting benefits from translation-invariant global descriptors, position estimation must preserve spatial information throughout the network [9], [10]. Consequently, the global average pooling layers used in the Classification-NN are replaced by flattening operations, preserving the spatial arrangement of the learned features (See Fig. D1). Separate convolutional branches extract these features from the intensity and autocorrelation measurements before these representations are fused using the same learnable photon-budget-dependent gating mechanism introduced in the previous section. The fused feature vector is then processed by a fully connected regression head that simultaneously predicts the positions and the emission probabilities of the emitters. For more details on the network and their training, please refer to Appendix D.

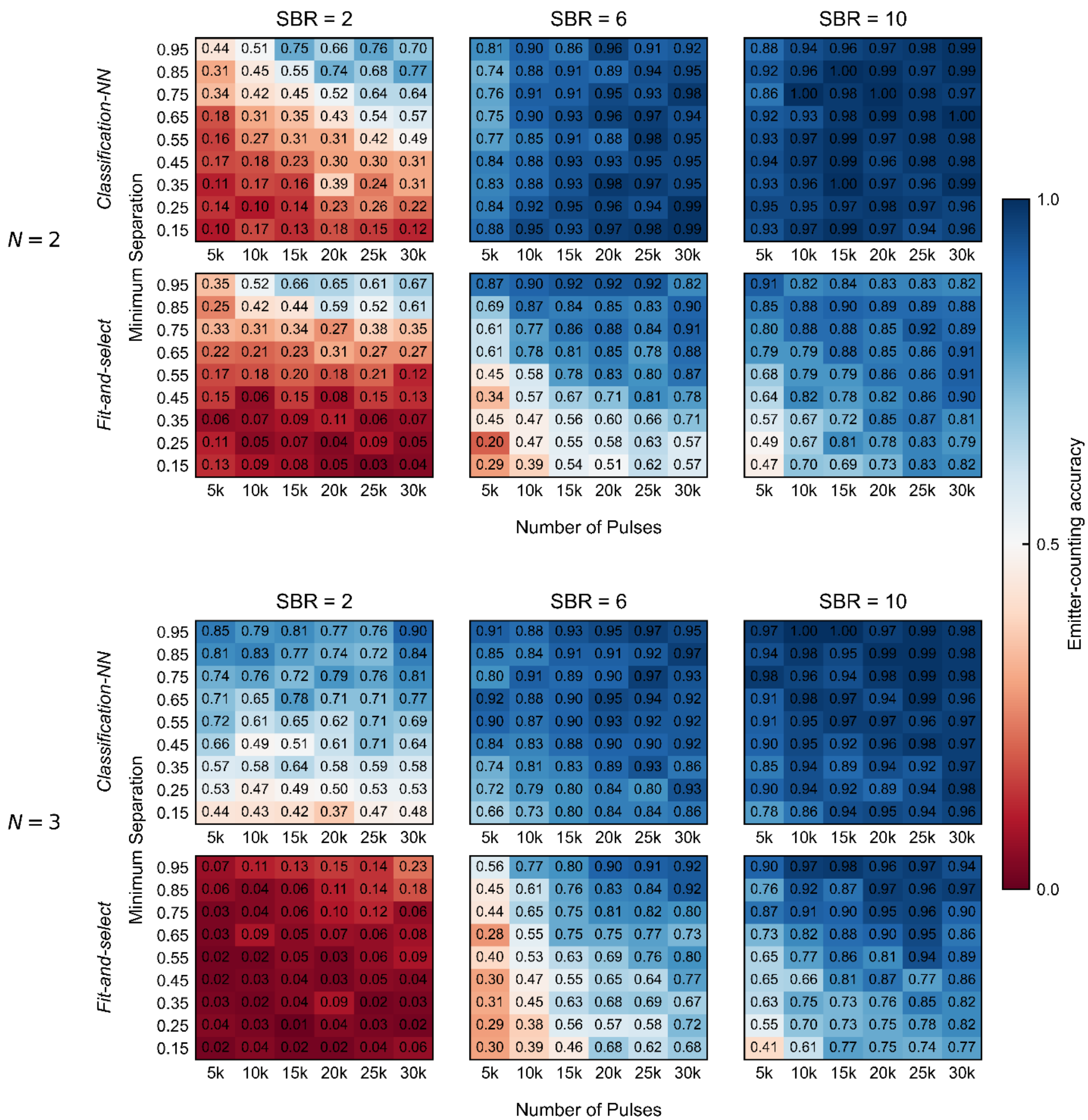

**Figure 4. Emitter-counting accuracy as a function of emitter separation and photon budget.** Results are shown for ground-truth emitter counts $N_{\mathrm{GT}} = 2$ and $N_{\mathrm{GT}} = 3$ (sections) and SBR = 2, 6 and 10 (columns). Within each section, the upper row shows results for the proposed Classification-NN, and the lower row shows results for the fit-and-select approach. Each heatmap shows emitter-counting accuracy as a function of the minimum ground-truth emitter separation and the photon budget $n_{\mathrm{pulse}}$, evaluated over Monte Carlo realizations on identical simulated datasets.

Three instances of the Regression-NN were trained separately for each $N = \{1,2,3\}$, with a dedicated model for each. The emitter separation was sampled uniformly in the range$[0.1,1.0]$. For analysis, the samples were grouped into bins of width 0.1 based on the ground-truth separation, e.g., separations in $[0.1, 0.2)$, $[0.2, 0.3)$, etc. are labelled with the mid-point of the range. Each bin had 100 samples per $n_{\text{pulse}}$. As in the previous section, we choose SBR = 2, 6 and 10 as our case studies.

Fig. 5 shows the comparison of position estimation performance between the fit-and-select approach and the Regression-NN, both conditioned on correct prior classification. We sweep across a range of photon budgets, ground-truth emitter separations, i.e., $\min_{i \neq j} \left| x_{\text{GT}}^{i} - x_{\text{GT}}^{j} \right|$ and SBRs. The first point to note is that the NN substantially outperforms the fit-and-select approach across all parameter regimes with more than a tenfold improvement in position-estimation precision at low separation regimes below 0.3, where standard parameter-fitting of the fit-and-select approach struggles to discern the closely spaced emitters. The Regression-NN, on the other hand, remains relatively stable across all emitter separations. A similar trend is seen for the fit-and-select approach's dependence on photon budgets. However, the fit-and-select approach improves with increasing photon budget only in the mid- to high-separation regime (>0.45), with lower separation regimes not showing any improvement even with maximal photon budget. The Regression-NN, on the other hand, shows a consistent improvement in response to increasing photon budgets. The SBR = 2 case is challenging for both models. The fit-and-select approach shows a constant poor performance across photon budgets and separations for $N_{\text{GT}} = 2$ and a very slight improvement with increasing separation for $N_{\text{GT}} = 3$. On the other hand, Regression-NN has substantially better position-estimation precision for both ground-truth emitter numbers.

As SBR decreases, the Regression-NN's performance degrades rather modestly, whereas the fit-and-select approach shows a sharper degradation. This behavior is consistent with the position-fitting problem becoming less well constrained as the background increases relative to the emitter signal and as overlapping PSFs make different position and emission probability configurations produce increasingly similar spatial profiles. The Regression-NN, in contrast, is trained directly for position estimation over a large ensemble of noisy realizations and can learn nonlinear, spatially preserving representations of $I(x)$ and $g^{(2)}(x)$ that are combined adaptively across photon budgets. These differences provide plausible explanations for its greater robustness, although this comparison does not isolate which component is responsible for the performance gain.

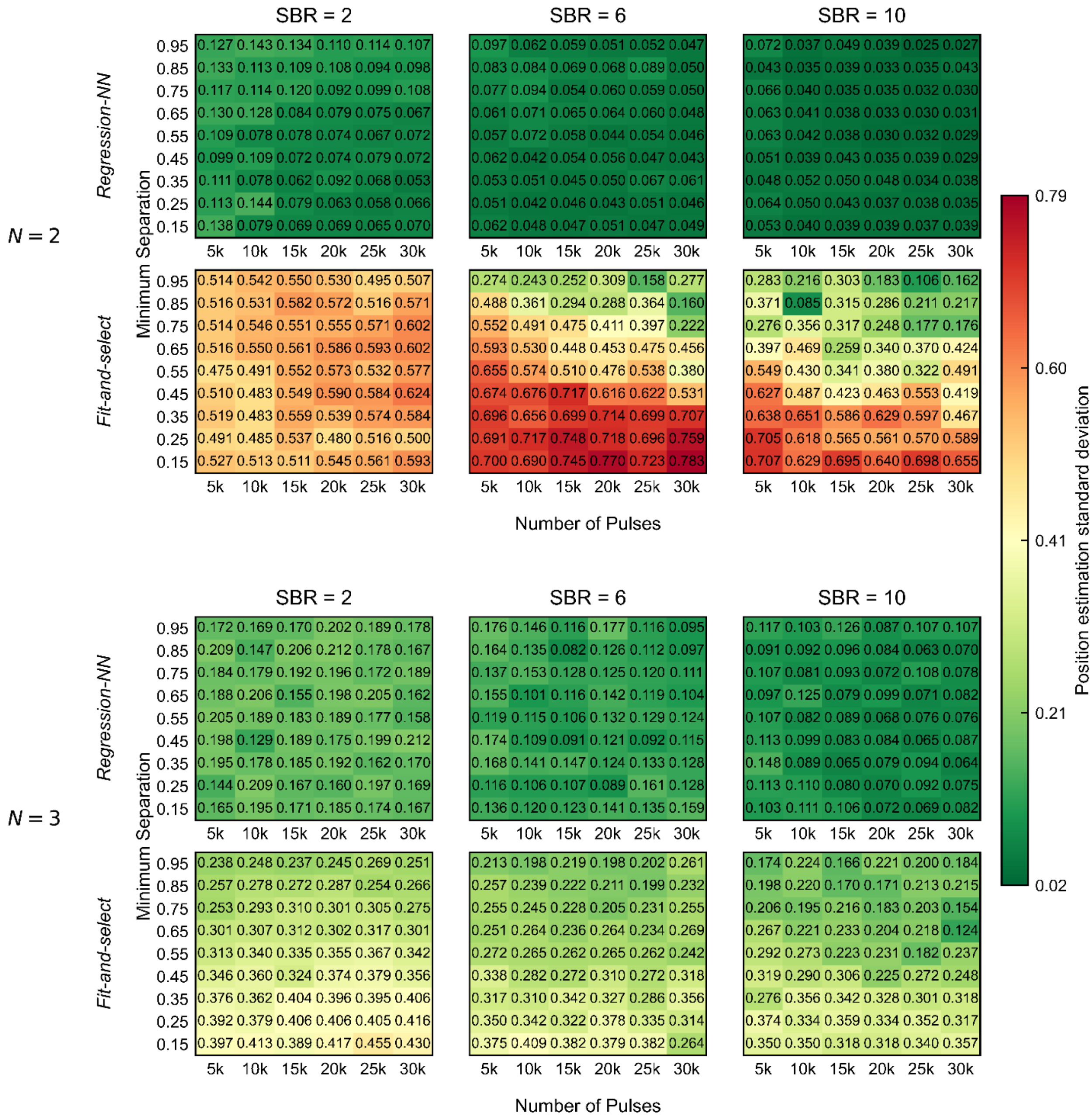

**Figure 5: Position-estimation standard deviation as a function of emitter separation and photon budget.** Results are shown for ground-truth emitter counts $N_{GT}$ = 2 and $N_{GT}$ = 3 (sections) and SBR 2, 6 and 10 (columns). Within each section, the upper row shows results for the proposed Regression-NN, and the lower row shows results for the existing fit-and-select approach. $N$ is fixed to the ground-truth value for both approaches. Each heatmap shows the standard deviation of the position-estimation error as a function of the minimum ground-truth emitter separation and the number of excitation pulses, evaluated over independent Monte Carlo realizations on identical simulated datasets.

We also note a peculiar behavior of the fit-and-select method. For SBR = 2 and $N_{\mathrm{GT}} = 2$, the precision is strongly uniform and equal to about 0.5. Furthermore, at higher SBR the precision becomes better in the high separation regime, but worse in the low separation regime. The plateau in the SBR = 2 case arises from the guess positions that land rather randomly throughout the measurement volume, including the measurement boundary. At SBR = 6 and 10, some of these estimates move closer to the ground truth positions due to improved position estimation while others still remain close at the measurement boundary. This arguably constitutes an improvement compared to the SBR = 2 case. However, the precision is computed from the standard deviation of errors, and thus becomes worse.

**Results: Localization under finite measurement-window truncation**

Having established that, in isolation, the learned inference methods outperform the fit-and-select approach in their task-specific metrics, we now assess the robustness of the two approaches to a separate source of information loss through truncation arising from the finite spatial extent of the measurement window. In our simulations, as in practical measurements, the spatial data for emitters located close to the edge of the sampled region will be severely truncated, reducing the information available for inference. To isolate the effect of this truncation from changes in emitter separation or background level, we consider two-emitter configurations that differ only in their position relative to the map's center.

We simulate $N_{GT}$ = 2 emitters with a fixed separation of $0.5$, relative brightness ratio $0.5$, and SBR $= 6$. The measurement grid spans $x \in [-2,2]$. Across the three configurations A (center around $x = 0$), B (centered around $x = 0.75$) and C (centered around $x = 1.5$) shown in Fig. 6, the emitter pair is progressively translated from the center of the measurement window toward the right boundary. For each configuration, the ground-truth emitter positions are fixed across independent realizations, such that the only variation between samples arises from photon statistics. As the emitters approach the boundary, an increasing fraction of the corresponding $I(x)$ and $g^{(2)}(x)$ profiles lie outside the sampled region, producing progressively stronger measurement truncation (Fig. 6, top row).

We plot the position-estimation standard deviation and the corresponding emitter-counting accuracy separately in the bottom row for the three configurations as a function of the photon budget. When the emitter-counting accuracy is below the $90\%$ threshold (used here as a criterion for practically reliable emitter counting), the corresponding position-estimation standard deviation is colored in gray, signifying unreliable localization.

As shown in Fig. 6, the effect of truncation changes not only the magnitude of the performance difference between the two approaches, but also the inference task in which the principal advantage arises. For configuration A, the Classification-NN and fit-and-select approaches achieve similar emitter-counting accuracy over most of the explored photon-budget range, mostly exceeding 90%. However, the main advantage of the NN framework is revealed in position estimation, for which the Regression-NN consistently exhibits a threefold lower position-estimation standard deviation. A similar behavior is observed for the configuration B, although the fit-and-select approach requires a somewhat larger photon budget, ~ 15,000 pulses, to reach reliable counting. Both approaches eventually achieve comparable emitter-counting accuracy, but the Regression-NN retains an over three-fold advantage in position-estimation precision.

The comparison changes markedly in the configuration C. Here, a significant fraction of the spatial data is truncated by the measurement boundary. The Classification-NN continues to maintain over 90% accuracy, whereas the fit-and-select approach is far from this threshold even at 30,000 pulses. Although a position-estimation standard deviation can still be evaluated for the fit-and-select estimator, its practical relevance is limited when the corresponding emitter count is not reliably recovered.

Taken together, these results show that the advantage of task-specific learned inference persists under a qualitatively different form of information degradation from those considered previously. When the spatial profiles are well sampled, the primary gain is improved position-estimation precision. As measurement truncation becomes severe, reliable emitter counting instead becomes the limiting step for the fit-and-select approach, while the neural networks remain robust under all such degradations.

## Discussion

The fit-and-select approach successfully exploits quantum correlations. However, it couples the two different tasks of emitter counting and position estimation by determining the emitter number through fits of the spatial parameters for each candidate model. As demonstrated in this work, this creates a vulnerability in emitter-count estimation, since errors or ambiguity in the multi-parameter spatial fits can propagate into the subsequent selection of $N$. The learned inference framework explicitly separates the two tasks and equips each task with a dedicated neural network that learns task-specific combinations of $I(x)$ and $g^{(2)}(x)$. The Classification-NN improves emitter-counting accuracy by inferring $N$ directly from the measured data rather than through selection among competing spatial fits. Independently from this, when the correct emitter count is known, the learned Regression-NN exhibits a substantially better position-estimation precision than the parametric fit-and-select approach. The learned inference framework outperforms the fit-and-select approach under photon-limited conditions, intense background, strongly overlapping emitter profiles and spatial data truncation.

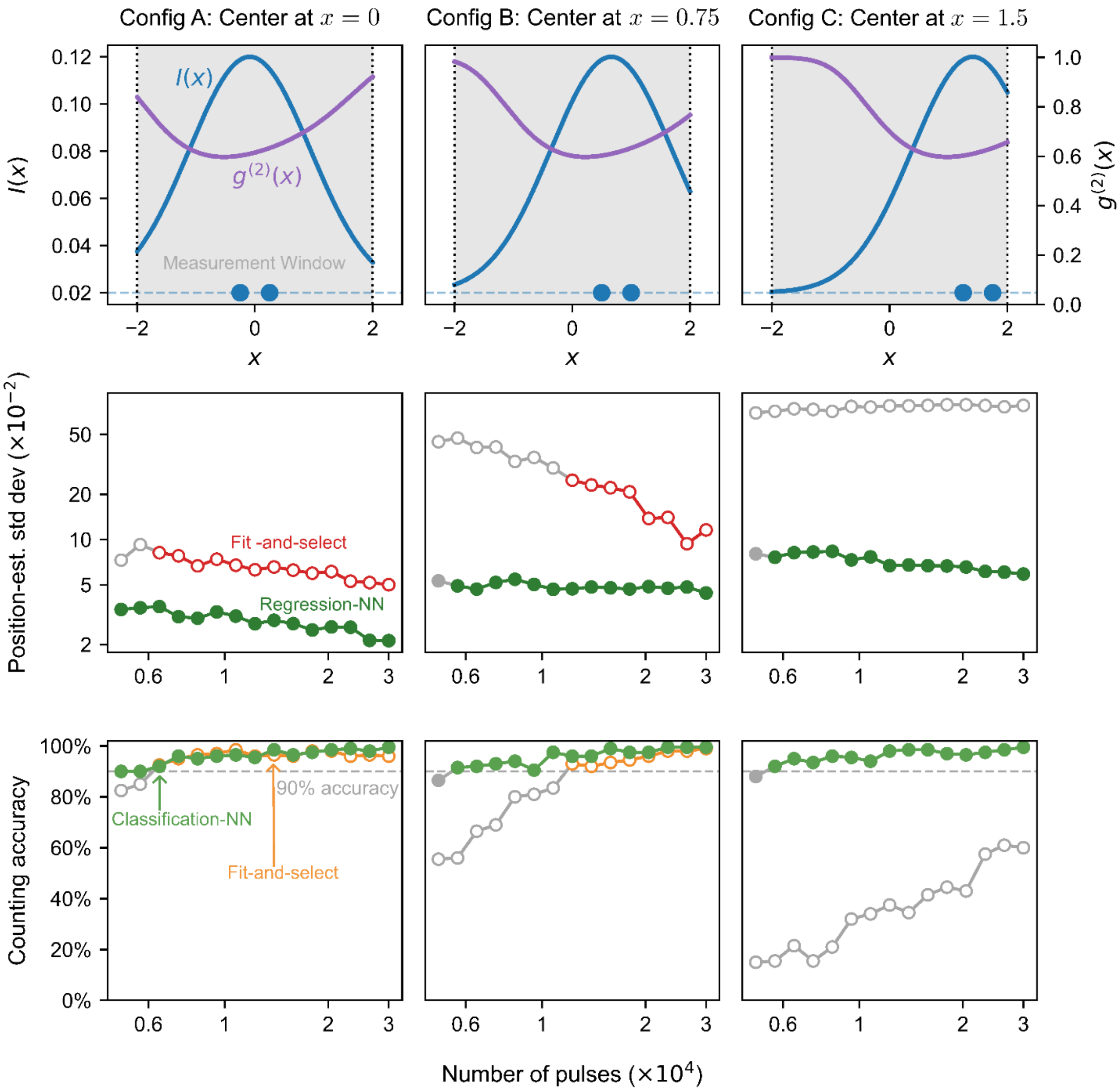

**Figure 6. Localization performance under finite measurement-window truncation at SBR = 6.** (a) Theoretical intensity, $I(x)$, and second-order correlation, $g^{(2)}(x)$, for three fixed two-emitter configurations with separation 0.5 and relative brightness ratio 0.5. From left to right, the emitter pair is translated from the center of the measurement region toward the right boundary. The shaded region between the dotted vertical lines indicates the measurement window, $x \in [-2,2]$. (b) Position-estimation standard deviation as a function of the number of excitation pulses for the corresponding configurations. The Regression-NN (green) and fit-and-select approach (red) are evaluated over all realizations. Gray markers indicate photon budgets for which the corresponding emitter-counting accuracy is below 90%. (c) Emitter-counting accuracy for the same configurations and photon budgets, shown for the Classification-NN (light green) and fit-and-select approach (orange). The dashed horizontal line indicates the 90% accuracy criterion.

The proposed framework is particularly well suited to applications where photon efficiency is critical, including biological imaging, single-particle tracking, and quantum sensing [32], [33], [34], [35]. In these settings, reducing the required photon budget can help mitigate photobleaching and phototoxicity while reducing acquisition time and enabling faster measurements [10], [36]. The robustness observed under finite measurement-window truncation may also be advantageous when the sampled field of view is small, i.e. the data acquisition is performed on less than 100 spatial pixels.

Several natural extensions follow from this work. The learned inference framework could, in principle, be extended from scanned microscopy to imaging arrays. SPAD-array detectors have already been incorporated into confocal microscopy to provide spatially resolved single-photon measurements, including for fluorescence fluctuation spectroscopy and single-particle tracking [33], [37], [38]. An extension of the present framework to such measurements could enable scan-less localization [22], [31], [39]. Although the present study considers one-dimensional localization of up to three emitters, the proposed framework is readily extendable to two- and three-dimensional imaging and to larger emitter populations through appropriate modifications to the network architecture and training data [10], [27], [29]. This would require learning from higher-dimensional spatial distributions and resolving a correspondingly larger set of spatial configurations. Incorporating experimentally relevant effects, including detector jitter, optical aberrations, and non-ideal photon statistics, will be essential for deployment in practical imaging systems. Finally, experimental validation on real photon-correlation measurements represents an important next step toward integrating the proposed framework into quantum-enhanced microscopy platforms.

Several limitations should be considered when interpreting these results. First, the study is based entirely on simulated one-dimensional measurements generated using a Gaussian point-spread function and idealized statistical models for the emitters and background. Consequently, the demonstrated performance does not establish robustness to experimentally specific effects such as detector jitter, optical aberrations, or deviations from the assumed photon statistics. Additional ambiguities may arise if more than three emitters are present in the focal volume. Finally, because the inference models are trained on simulated data, their performance depends on the agreement between the training distribution and the statistics of the target experimental system. Distributional differences between simulation and experiment could therefore reduce localization performance [9], [10], [32].

**Acknowledgements:**

This material is based upon work supported by the U.S. Department of Energy, Office of Science, Office of Biological and Environmental Research, under Award Number DE-SC0023167.

Disclaimer: This report was prepared as an account of work sponsored by an agency of the United States Government. Neither the United States Government nor any agency thereof, nor any of their employees, makes any warranty, express or implied, or assumes any legal liability or responsibility for the accuracy, completeness, or usefulness of any information, apparatus, product, or process disclosed, or represents that its use would not infringe privately owned rights. Reference herein to any specific commercial product, process, or service by trade name, trademark, manufacturer, or otherwise does not necessarily constitute or imply its endorsement, recommendation, or favoring by the United States Government or any agency thereof. The views and opinions of authors expressed herein do not necessarily state or reflect those of the United States Government or any agency thereof.

## Appendix A. Physical and Statistical Model

We consider a one-dimensional confocal scanning geometry containing $N_{\mathrm{GT}}$ single-photon emitters (SPEs) located at positions $\{x_{\mathrm{GT}}^{i}\}_{i=1}^{N_{\mathrm{GT}}}$within the confocal volume. Each emitter produces antibunched photoluminescence under pulsed optical excitation and is embedded in a spatially uniform classical background.

### Optical excitation and emission model

The system is driven by a sequence of $n_{\mathrm{pulse}}$identical optical pump pulses. The pulse duration is assumed to be much shorter than the excited-state lifetime $\tau_{\mathrm{ES}}$, while the repetition period is much longer than $\tau_{\mathrm{ES}}$, ensuring independent emission events across pulses.

Each emitter produces at most one photon per pulse. Emitters are allowed to be non-identical in brightness (emission efficiency). The spatial emission profile of each emitter is modeled as a Gaussian point spread function (PSF):

$$I_{\mathrm{GT}}^{i}(x) = \frac{A_{\mathrm{GT}}^{i}}{\sqrt{2\pi}\sigma} e^{-\frac{\left(x - x_{\mathrm{GT}}^{i}\right)^2}{2\sigma^2}} \tag{2}$$

where $\sigma$is the PSF width (used as the unit of length throughout), and $A_{\mathrm{GT}}^{i}$is the peak brightness of emitter $i$ which is the product of emission, collection and detection efficiency. A spatially uniform background contributes Poisson-distributed photons with mean intensity $I_{\mathrm{BG}}$. The signal-to-background ratio (SBR) is defined as the ratio of the maximum total intensity to the background intensity.

**Peak Intensity Assignment**

The peak emission probabilities (or peak intensities) of individual emitters are determined by fixing the background intensity $I_{\mathrm{BG}}$, the signal-to-background ratio (SBR), and the relative intensity ratios between emitters.

The emission probabilities $A_{\mathrm{GT}}^{i}$ are determined such that:

1. The relative intensity ratios between emitters are fixed (e.g., $A_{\mathrm{GT}}^{2} = \alpha_{\mathrm{GT}} A_{\mathrm{GT}}^{1}$, $A_{\mathrm{GT}}^{3} = \beta_{\mathrm{GT}} A_{\mathrm{GT}}^{1}$),
2. The global maximum of the total intensity satisfies the SBR constraint:

$$\max_{x} I(x) = \mathrm{SBR} \times I_{\mathrm{BG}}.$$

To enforce this, we first determine the location $x_{\mathrm{peak}}$of the global maximum of $I(x)$ by solving $\frac{dI}{dx} = 0$. The Gaussian basis functions are then evaluated at $x_{\mathrm{peak}}$, and the emission probabilities are obtained from:

$$A_{\mathrm{GT}}^{1} = \frac{I_{\mathrm{BG}}(SBR - 1)\sqrt{2\pi}\sigma}{\sum_{i} r_{\mathrm{GT}}^{i} \exp\left(-\frac{\left(x_{\mathrm{peak}} - x_{\mathrm{GT}}^{i}\right)^{2}}{2\sigma^{2}}\right)} \tag{3}$$

where $r_{\mathrm{GT}}^{i}$are the relative intensity ratios (i.e., $1, \alpha_{\mathrm{GT}}$ or $\beta_{\mathrm{GT}}$). The remaining emission probabilities follow directly from the imposed ratios. This procedure ensures that the combined emitter signal achieves the desired SBR while preserving controlled relative brightness across emitters.

**Detection and observables**

The emission is collected by a Hanbury–Brown–Twiss (HBT) interferometer consisting of a 50:50 beamsplitter followed by two single-photon detectors. For each spatial scan position $x_k$, two observables are extracted from the detector click records. The intensity data are defined as:

$$I(x) = \langle n_1(x) + n_2(x) \rangle \quad (4)$$

where $n_1$ and $n_2$ denote the binary detection outcomes of detectors 1 and 2 for an individual excitation pulse, <.> denotes an average over the excitation pulses acquired at spatial position $x$.

The second-order autocorrelation data is defined as:

$$g^{(2)}(x) = \frac{\langle n_1(x) n_2(x) \rangle}{\langle n_1(x) \rangle \langle n_2(x) \rangle}, \quad (5)$$

where $\langle n_1(x) n_2(x) \rangle$ is the mean number of coincident detections per excitation pulse, with the coincidence defined as simultaneous detection events in the two detector channels. The denominator is the product of the mean detection probabilities of the two detectors over the same excitation pulses.

For perfectly antibunched emission, $g^{(2)}(0) = 0$. Background emission is characterized by $g^{(2)}(0) = 1$. The measured correlation signal therefore encodes local information about the number of neighboring emitters.

The scan consists of $n_x = 15$ spatial sampling points uniformly distributed over the interval $[-2\sigma, 2\sigma]$. The measured dataset is thus

$$\mathcal{D} = \{I(x_k), g^{(2)}(0; x_k)\}_{k=1}^{n_x}.$$

**Appendix B. Fit-and-Select Approach and Model Selection**

We implement a fit-and-select approach following the approach introduced by Vasquez-Lozano *et al.* [21]. In this approach, the emitter number and emitter positions are inferred simultaneously from the measured intensity and second-order correlation maps. For each hypothesized emitter number $N \in 1,2,3$, a parametric forward model $I_N(x; \theta_N)$ and $g_N^{(2)}(x; \theta_N)$ is constructed, where $\theta_N$ contains the emitter positions and emission probabilities.

For each emitter-number guess, the model parameters are estimated by minimizing the weighted least-squares objective:

$$\mathcal{L}_N(\theta_N) = (1-\lambda) \sum_x \left( \frac{I_N(x; \theta_N) - I_{\text{data}}(x)}{s_I} \right)^2 + \lambda \sum_x \left( \frac{g_N^{(2)}(x; \theta_N) - g_{\text{data}}^{(2)}(x)}{s_g} \right)^2 \quad (6)$$

In Eq. (6), $\lambda$ controls the relative contribution of the correlation measurement, and $s_I$ and $s_g$ denote the standard deviations of the measured intensity and correlation profiles, respectively. The fitted parameter vector for each emitter-number guess is obtained by numerical minimization: $\hat{\theta}_N = \underset{\theta_N}{\operatorname{argmin}} L_N(\theta_N)$. The minimization is performed using the L-BFGS-B algorithm [40] with fixed initialization of the emitter positions and emission probabilities. Position parameters are constrained to the scan range and emission probabilities to positive values. This procedure yields a fitted model for each guess of emitter number.

The fitted models are subsequently compared using the Akaike Information Criterion (AIC), which accounts for both the goodness of fit and the number of fitted parameters [25]. The AIC is calculated as:

$$\mathrm{AIC}_N = 2k_N + n_{\mathrm{data}} \ln\left(\frac{\mathcal{L}_N(\hat{\theta}_N)}{n_{\mathrm{data}}}\right) \tag{7}$$

In Eq. (7), $k_N$ is the number of fitted parameters and $n_{\mathrm{data}}$ is the total number of measured intensity and correlation values included in the fit. The guess with the minimum AIC is selected as the estimated emitter number, together with its corresponding fitted parameters.

We compute the position-estimation standard deviation of the error $E = \min_{\pi \in S_N} \frac{1}{N} \sum_{i=1}^{N} \left| x^{\pi(i)} - x_{\mathrm{GT}}^{i} \right|$. Here, $x^i$ is the predicted position, $x_{\mathrm{GT}}^{i}$ is the ground truth position and $S_N$ denotes the set of all permutations of $\{1, \dots, N\}$. The reported standard deviation is Std(E) across multiple independent runs. For position-estimation comparison, N is fixed to the ground-truth value, N = $N_{\mathrm{GT}}$.

## Appendix C. Classification Neural Network

Emitter counting is performed using a supervised dual-branch convolutional neural network (CNN).

### Input representation

Each sample consists of:

- an intensity vector $I(x_k) \in \mathbb{R}^{15}$,
- a zero-delay second-order correlation vector $g^{(2)}(x_k) \in \mathbb{R}^{15}$,

- the logarithm of the number of excitation pulses $\log_{10}(n_{\text{pulse}})$.

**Network architecture**

The CNN consists of two parallel one-dimensional convolutional branches: one processing intensity, and the other processing the correlation data.

Each branch consists of two one-dimensional convolutional layers with 16 and 32 filters (kernel size 3, ReLU activation), followed by global average pooling, producing 32-dimensional feature vectors $f_I$and $f_{g^{(2)}}$. A 32-dimensional gating vector conditioned on $\log_{10}(n_{\text{pulse}})$adaptively weights the two feature streams before fusion. The fused representation is passed through two fully connected layers with 64 and 32 neurons followed by a softmax output over $N \in \{1,2,3\}$.

This architecture allows the network to flexibly combine correlation and intensity features across photon budgets.

**Training method**

Training data were generated using Monte Carlo simulations of emitter systems with $N = 1$ to $3$ emitters. For each sample, the emitter count was randomly drawn from {1,2,3} with equal probability. Emitter positions were sampled uniformly within the range $[-2, 2]$, subject to a minimum separation constraint of 0.1 (in units of the point spread function width $\sigma$) to avoid degenerate configurations.

Relative emitter brightness was controlled via multiplicative ratios. For two-emitter systems, the ratio parameter $\alpha$ was sampled uniformly from $[0.1,1.0]$, such that $A^2 = \alpha A^1$. For three-emitter systems, an additional ratio $\beta \in [0.1, 1.0]$ was independently sampled, defining $A^3 = \beta A^1$. The absolute emission probabilities were then determined by enforcing a fixed signal-to-background ratio (SBR), as described earlier in Appendix A.

The background intensity was fixed at $I_{\text{BG}} = 0.02$, and datasets were generated at specified SBR values. Photon statistics were simulated by sampling the number of excitation pulses from the set {500, 1000, 3000, 5000, 10000, 20000, 30000}, enabling the model to learn across a wide range of signal-to-noise conditions.

For each configuration, intensity and second-order correlation $g^{(2)}(0)$ were computed over a fixed 15-point spatial grid spanning $[-2, 2]$. A total of $3.5 \times 10^5$ samples were generated for training.

The dataset was split into training (80%) and testing (20%) subsets using stratified sampling to preserve class balance across emitter counts. Input features were standardized using training-set statistics. To improve generalization, the training data was augmented using random circular shifts and mirror reflections of the spatial profiles. These operations preserve the underlying emitter count while exposing the network to different spatial realizations of identical physical configurations.

The model was trained using the Adam optimizer with a learning rate of $3 \times 10^{-4}$, categorical cross-entropy loss, and a batch size of 128. Training was performed for up to 100 epochs with early stopping based on validation accuracy (patience = 5), and the best-performing model was retained.

**Appendix D. Regression Neural Network**

Position estimation is performed using a supervised dual-branch convolutional neural network conditioned on the known emitter number. The Regression-NN directly estimates the emitter positions and relative emission probabilities from the measured intensity and correlation data.

**Input representation**

Each sample consists of

- an intensity vector $I \in \mathbb{R}^{15}$,
- a correlation vector $g^{(2)} \in \mathbb{R}^{15}$,
- the logarithm of the number of excitation pulses $\log_{10}(n_{\text{pulse}})$.

The network predicts the target vector $(x^1, A^1)$ for N=1, $(x^1, x^2, A^1, A^2)$ for $N = 2$ and $(x^1, x^2, x^3, A_1, A^2, A^3)$ for $N = 3$.

**Network architecture**

The network consists of two parallel one-dimensional convolutional branches: one processing intensity and one processing correlation. Each branch contains three one-dimensional convolutional layers with 32, 64, and 64 filters (kernel size 3, ReLU activation), followed by a Flatten operation in contrast to GAP used for Classification-NN (Fig. D1).

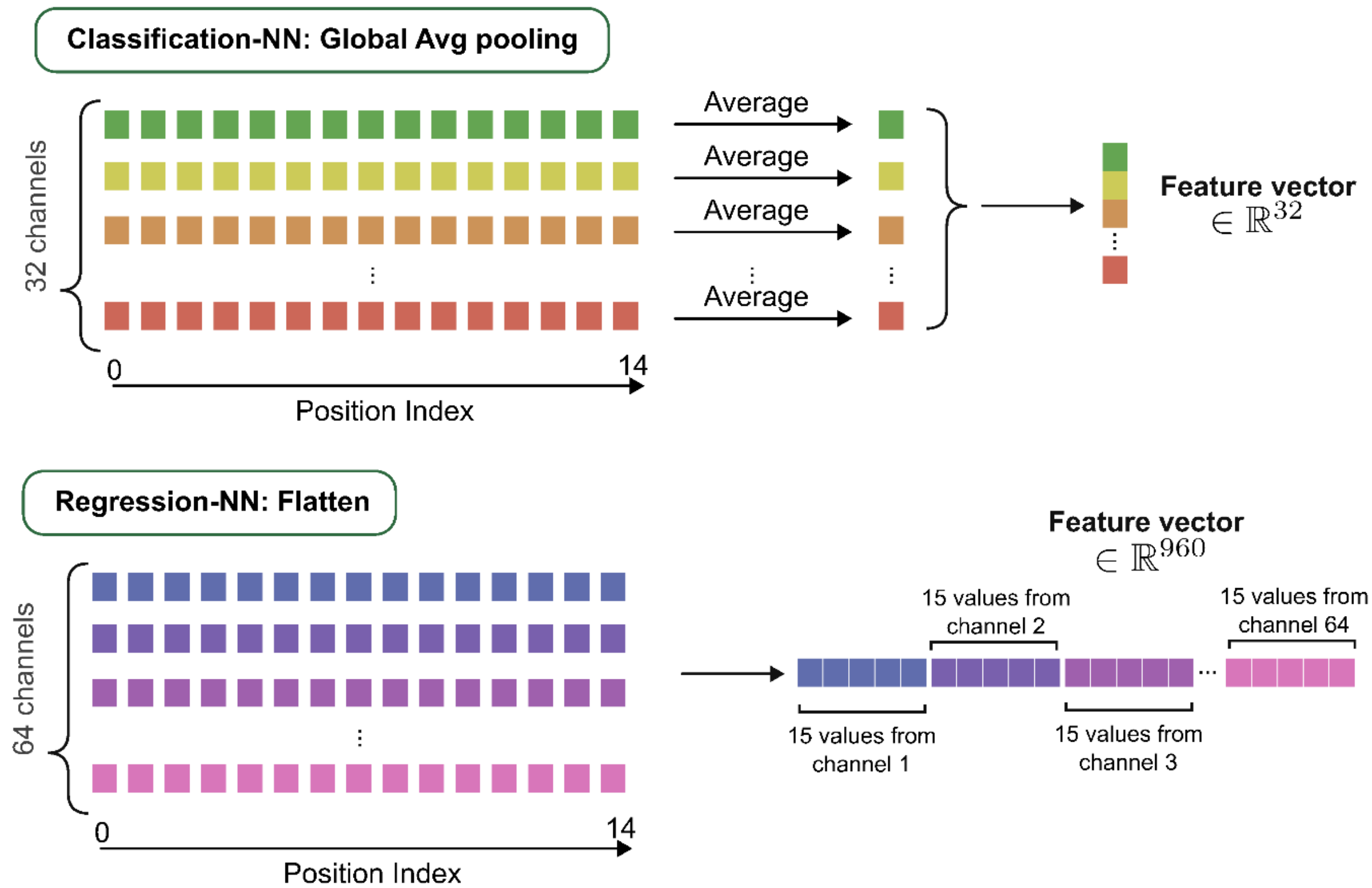

**Figure D1. Comparison of Global Average Pooling vs Flatten:** The Classification-NN applies global average pooling independently to each of the 32 CNN channels, averaging the 15 spatially indexed feature values to produce one value per channel and a 32-dimensional feature vector. In contrast, the Regression-NN flattens all 15 values from each of its 64 channels, producing a 960-dimensional feature vector.

The resulting feature vectors are fused using a learnable gating vector conditioned on $\log_{10}(n_{\text{pulse}})$, which adaptively weights the contributions of the intensity and correlation branches on a channel-by-channel basis.

The fused representation is passed through a fully connected regression network comprising dense layers with 256, 128, and 64 neurons, followed by a linear output layer that jointly predicts the emitter positions and emission probabilities.

### Training method

Training data were generated using Monte Carlo simulations separately for different numbers of emitters. Emitter positions were sampled uniformly while maintaining a minimum separation of $0.1$. Relative emitter brightnesses were randomized by sampling the brightness ratio $\alpha$ uniformly from [0.1,1], with the absolute probability subsequently determined from the prescribed signal-to-background ratio as described previously.

Photon statistics were simulated by randomly sampling the number of excitation pulses from $\{500, 1000, 3000, 5000, 10000, 20000, 30000, 50000\}$. Intensity and second-order correlation measurements were computed over the fixed 15-point spatial grid spanning $[-2,2]$. A total of $3.5 \times 10^5$ samples were generated.

The dataset was randomly divided into training (80%) and testing (20%) subsets. Prior to training, the input features were standardized using statistics computed from the training set. To exploit the mirror symmetry of the position-estimation problem, the training data were augmented using physically valid spatial reflections, with the corresponding emitter positions and emission probabilities transformed accordingly.

The model was trained using the Adam optimizer with a learning rate of $10^{-3}$, mean squared error loss, and a batch size of 128. Training was performed for up to 100 epochs with early stopping based on the validation loss (patience = 10), and the model parameters corresponding to the lowest validation loss were retained.

**Bibliography**

[1] M. J. Rust, M. Bates, and X. Zhuang, “Sub-diffraction-limit imaging by stochastic optical reconstruction microscopy (STORM),” *Nat. Methods*, vol. 3, no. 10, pp. 793–796, Oct. 2006, doi: 10.1038/nmeth929.

[2] E. Betzig *et al.*, “Imaging Intracellular Fluorescent Proteins at Nanometer Resolution,” *Science*, vol. 313, no. 5793, pp. 1642–1645, Sep. 2006, doi: 10.1126/science.1127344.

[3] S. T. Hess, T. P. K. Girirajan, and M. D. Mason, “Ultra-High Resolution Imaging by Fluorescence Photoactivation Localization Microscopy,” *Biophys. J.*, vol. 91, no. 11, pp. 4258–4272, Dec. 2006, doi: 10.1529/biophysj.106.091116.

[4] R. E. Thompson, D. R. Larson, and W. W. Webb, “Precise Nanometer Localization Analysis for Individual Fluorescent Probes,” *Biophys. J.*, vol. 82, no. 5, pp. 2775–2783, May 2002, doi: 10.1016/S0006-3495(02)75618-X.

[5] K. I. Mortensen, L. S. Churchman, J. A. Spudich, and H. Flyvbjerg, “Optimized localization analysis for single-molecule tracking and super-resolution microscopy,” *Nat. Methods*, vol. 7, no. 5, pp. 377–381, May 2010, doi: 10.1038/nmeth.1447.

[6] K. Hermon, S. Schidorsky, Y. Razvag, O. Yakovian, and E. Sherman, “Time-correlated single molecule localization microscopy enhances resolution and fidelity,” *Sci. Rep.*, vol. 10, no. 1, p. 16212, Oct. 2020, doi: 10.1038/s41598-020-72812-y.

[7] R. J. Ober, S. Ram, and E. S. Ward, “Localization Accuracy in Single-Molecule Microscopy,” *Biophys. J.*, vol. 86, no. 2, pp. 1185–1200, Feb. 2004, doi: 10.1016/S0006-3495(04)74193-4.

[8] S. Ram, E. S. Ward, and R. J. Ober, “Beyond Rayleigh’s criterion: A resolution measure with application to single-molecule microscopy,” *Proc. Natl. Acad. Sci.*, vol. 103, no. 12, pp. 4457–4462, Mar. 2006, doi: 10.1073/pnas.0508047103.

[9] E. Nehme, L. E. Weiss, T. Michaeli, and Y. Shechtman, “Deep-STORM: super-resolution single-molecule microscopy by deep learning,” *Optica*, vol. 5, no. 4, p. 458, Apr. 2018, doi: 10.1364/OPTICA.5.000458.

[10] A. Speiser *et al.*, “Deep learning enables fast and dense single-molecule localization with high accuracy,” *Nat. Methods*, vol. 18, no. 9, pp. 1082–1090, Sep. 2021, doi: 10.1038/s41592-021-01236-x.

[11] H. Babcock, Y. M. Sigal, and X. Zhuang, “A high-density 3D localization algorithm for stochastic optical reconstruction microscopy,” *Opt. Nanoscopy*, vol. 1, no. 1, p. 6, 2012, doi: 10.1186/2192-2853-1-6.

[12] O. Mandula, I. Š. Šestak, R. Heintzmann, and C. K. I. Williams, “Localisation microscopy with quantum dots using non-negative matrix factorisation,” *Opt. Express*, vol. 22, no. 20, p. 24594, Oct. 2014, doi: 10.1364/OE.22.024594.

[13] M. Tsang, R. Nair, and X.-M. Lu, “Quantum Theory of Superresolution for Two Incoherent Optical Point Sources,” *Phys. Rev. X*, vol. 6, no. 3, p. 031033, Aug. 2016, doi: 10.1103/PhysRevX.6.031033.

[14] C. S. Smith, N. Joseph, B. Rieger, and K. A. Lidke, “Fast, single-molecule localization that achieves theoretically minimum uncertainty,” *Nat. Methods*, vol. 7, no. 5, pp. 373–375, May 2010, doi: 10.1038/nmeth.1449.

[15] S. J. Holden, S. Uphoff, and A. N. Kapanidis, "DAOSTORM: an algorithm for high-density super-resolution microscopy," *Nat. Methods*, vol. 8, no. 4, pp. 279–280, Apr. 2011, doi: 10.1038/nmeth0411-279.

[16] L. Zhu, W. Zhang, D. Elnatan, and B. Huang, "Faster STORM using compressed sensing," *Nat. Methods*, vol. 9, no. 7, pp. 721–723, Jul. 2012, doi: 10.1038/nmeth.1978.

[17] O. Schwartz and D. Oron, "Improved resolution in fluorescence microscopy using quantum correlations," *Phys. Rev. A*, vol. 85, no. 3, p. 033812, Mar. 2012, doi: 10.1103/PhysRevA.85.033812.

[18] D. Gatto Monticone *et al.*, "Beating the Abbe Diffraction Limit in Confocal Microscopy via Nonclassical Photon Statistics," *Phys. Rev. Lett.*, vol. 113, no. 14, p. 143602, Sep. 2014, doi: 10.1103/PhysRevLett.113.143602.

[19] H. J. Kimble, M. Dagenais, and L. Mandel, "Photon Antibunching in Resonance Fluorescence," *Phys. Rev. Lett.*, vol. 39, no. 11, pp. 691–695, Sep. 1977, doi: 10.1103/PhysRevLett.39.691.

[20] J. G. Worboys, D. W. Drumm, and A. D. Greentree, "Quantum multilateration: Subdiffraction emitter pair localization via three spatially separate Hanbury Brown and Twiss measurements," *Phys. Rev. A*, vol. 101, no. 1, p. 013810, Jan. 2020, doi: 10.1103/PhysRevA.101.013810.

[21] J. J. Vasquez-Lozano, S. Li, and A. D. Greentree, "Scanning quantum correlation microscopy with few emitters," *New J. Phys.*, vol. 26, no. 4, p. 043001, Apr. 2024, doi: 10.1088/1367-2630/ad31d5.

[22] Y. Israel, R. Tenne, D. Oron, and Y. Silberberg, "Quantum correlation enhanced super-resolution localization microscopy enabled by a fibre bundle camera," *Nat. Commun.*, vol. 8, no. 1, p. 14786, Mar. 2017, doi: 10.1038/ncomms14786.

[23] K. D. Weston, M. Dyck, P. Tinnefeld, C. Müller, D. P. Herten, and M. Sauer, "Measuring the Number of Independent Emitters in Single-Molecule Fluorescence Images and Trajectories Using Coincident Photons," *Anal. Chem.*, vol. 74, no. 20, pp. 5342–5349, Oct. 2002, doi: 10.1021/ac025730z.

[24] F. Huang, S. L. Schwartz, J. M. Byars, and K. A. Lidke, "Simultaneous multiple-emitter fitting for single molecule super-resolution imaging," *Biomed. Opt. Express*, vol. 2, no. 5, p. 1377, May 2011, doi: 10.1364/BOE.2.001377.

[25] H. Akaike, "A new look at the statistical model identification," *IEEE Trans. Autom. Control*, vol. 19, no. 6, pp. 716–723, Dec. 1974, doi: 10.1109/TAC.1974.1100705.

[26] M. Aßmann, "Quantum-Optically Enhanced STORM (QUEST) for Multi-Emitter Localization," *Sci. Rep.*, vol. 8, no. 1, p. 7829, May 2018, doi: 10.1038/s41598-018-26271-1.

[27] S. Li *et al.*, "Localising two sub-diffraction emitters in 3D using quantum correlation microscopy," *New J. Phys.*, vol. 26, no. 3, p. 033036, Mar. 2024, doi: 10.1088/1367-2630/ad31d4.

[28] R. A. Bartels, G. Murray, J. Field, and J. Squier, "Super-Resolution Imaging by Computationally Fusing Quantum and Classical Optical Information," *Intell. Comput.*, vol. 2022, p. 0003, Jan. 2022, doi: 10.34133/icomputing.0003.

[29] S. Li, W. Li, V. V. Yakovlev, A. Kealy, and A. D. Greentree, “En route to nanoscopic quantum optical imaging: counting emitters with photon-number-resolving detectors,” *Opt. Express*, vol. 30, no. 8, p. 12495, Apr. 2022, doi: 10.1364/OE.454412.

[30] D. Y. M. Peng *et al.*, “All-optical determination of one or two emitters using polarization-enhanced photon correlations of nitrogen-vacancy centers in diamond,” *Phys. Rev. A*, vol. 109, no. 2, p. 023723, Feb. 2024, doi: 10.1103/PhysRevA.109.023723.

[31] C. Seitz, C. Evans-Molina, and J. Liu, “Counting fluorescent emitters with a single photon avalanche diode array,” *Sci. Rep.*, Aug. 2026, doi: 10.1038/s41598-026-66870-x.

[32] Z. A. Kudyshev *et al.*, “Machine learning assisted quantum super-resolution microscopy,” *Nat. Commun.*, vol. 14, no. 1, p. 4828, Aug. 2023, doi: 10.1038/s41467-023-40506-4.

[33] A. Bucci *et al.*, “4D Single-particle tracking with asynchronous read-out single-photon avalanche diode array detector,” *Nat. Commun.*, vol. 15, no. 1, p. 6188, Jul. 2024, doi: 10.1038/s41467-024-50512-9.

[34] R. Tenne *et al.*, “Super-resolution enhancement by quantum image scanning microscopy,” *Nat. Photonics*, vol. 13, no. 2, pp. 116–122, Feb. 2019, doi: 10.1038/s41566-018-0324-z.

[35] F. Balzarotti *et al.*, “Nanometer resolution imaging and tracking of fluorescent molecules with minimal photon fluxes,” *Science*, vol. 355, no. 6325, pp. 606–612, Feb. 2017, doi: 10.1126/science.aak9913.

[36] J. Icha, M. Weber, J. C. Waters, and C. Norden, “Phototoxicity in live fluorescence microscopy, and how to avoid it,” *BioEssays*, vol. 39, no. 8, p. 1700003, Aug. 2017, doi: 10.1002/bies.201700003.

[37] C. Bruschini, H. Homulle, I. M. Antolovic, S. Burri, and E. Charbon, “Single-photon avalanche diode imagers in biophotonics: review and outlook,” *Light Sci. Appl.*, vol. 8, no. 1, p. 87, Sep. 2019, doi: 10.1038/s41377-019-0191-5.

[38] E. Slenders *et al.*, “Confocal-based fluorescence fluctuation spectroscopy with a SPAD array detector,” *Light Sci. Appl.*, vol. 10, no. 1, p. 31, Feb. 2021, doi: 10.1038/s41377-021-00475-z.

[39] J. J. Vasquez-Lozano, Q. Sun, S. Li, and A. D. Greentree, “Quantifying the advantage of photon correlation microscopy using arrays of single-photon detectors,” *Opt. Express*, vol. 33, no. 21, p. 44858, Oct. 2025, doi: 10.1364/OE.568719.

[40] R. H. Byrd, P. Lu, J. Nocedal, and C. Zhu, “A Limited Memory Algorithm for Bound Constrained Optimization,” *SIAM J. Sci. Comput.*, vol. 16, no. 5, pp. 1190–1208, Sep. 1995, doi: 10.1137/0916069.